\documentstyle[aps,prl]{revtex}
\begin{document}
\bibliographystyle{unsrt}

 \newcommand{\um}[1]{\"{#1}}
 \newcommand{\uck}[1]{\o}
 \renewcommand{\Im}{{\protect\rm Im}}
 \renewcommand{\Re}{{\protect\rm Re}}
 \newcommand{\ket}[1]{\mbox{$|#1\protect\rangle$}}
 \newcommand{\bra}[1]{\mbox{$\protect\langle#1|$}}
 \newcommand{\proj}[1]{\mbox{$\ket{#1}\bra{#1}$}}
 \newcommand{\expect}[1]{\mbox{$\protect\langle #1 \protect\rangle$}}
 \newcommand{\inner}[2]{\mbox{$\protect\langle #1 | #2
\protect\rangle$}}

\title{An atom optics experiment to investigate faster-than-light
tunneling}

\author{A.M. Steinberg, S. Myrskog, Han Seb Moon$^*$, Hyun Ah Kim$^*$, Jalani 
Fox, and Jung Bog Kim$^*$}

\address{Department of Physics, University of Toronto \\
Toronto, ONT M5S 1A7 CANADA\\ 
$^{*}$ Permanent address: Dept. of Physics, Korea Nat. 
Univ. of Education,\\ 
Chungbuk, Korea 363-791}

\date{\today}
\maketitle

\begin{abstract}
We describe a series of atom optics experiments underway at Toronto 
for investigating tunnelling interaction times of various sorts.
We begin by discussing some outstanding issues and confusions related
to the question of whether or not superluminal tunnelling can be
construed as true faster-than-light ``signal propagation,'' a question
which we answer in the negative.  We then argue that atom optics is
an arena ideally suited for addressing a variety of remaining
questions about how, where, and for how long a particle interacts with
a tunnel barrier.  We present recent results on a modified 
``delta-kick cooling'' scheme which we have used to prepare Rubidium
atoms with one-dimensional de Broglie wavelengths on the order of
an optical wavelength, along with simulations showing that from these
temperatures, we will be able to use acousto-optically modulated
dipole-force barriers to velocity-select ultracold atom samples ideal
for future tunnelling experiments.
\end{abstract}

 \vskip1pc

\section{Introduction: Superluminal tunnelling and information}

How long does it take a particle to tunnel through a barrier?
This question, first raised by 1932 
\cite{Condon=1931,MacColl=1932,Wigner=1955}, still arouses
extreme controversy today (see, e.g., 
\cite{Nimtz=1994,Steinberg=1994COM}).  
The subject went through a
renaissance theoretically in the 
1980s\cite{Buttiker=1982,Hauge=1989,Landauer=1994}, followed
by much experimental work in the 1990s, particularly using
electromagnetic analogs to tunnelling such as single-photon
transmission through ``photonic bandgap'' 
media\cite{Steinberg=1993PRL} and microwave transmission through
waveguides beyond cutoff\cite{Enders=1993}.  We will not review
our single-photon experiments here, as they have been extensively
discussed elsewhere.  In particular, we refer the reader to a
recent review\cite{Chiao=1997ProgOpt}, in which we discuss the history
of the tunnelling-time conundrum, along with the experiments on
electromagnetism and the thorny issues of superluminality and
causality.

For what the experiments have confirmed is the validity of the
stationary-phase approximation for wave packet propagation through
a tunnel barrier (a time frequently referred to as the ``phase time,''
but which we prefer to term the ``group delay'' in order to avoid
misunderstandings about phase vs. group velocities).  
At the heart of the problem is the fact that the 
kinetic energy of a particle inside a tunnel barrier is negative, so 
that a semiclassical estimate of its velocity becomes imaginary.  This 
makes it impossible to make the na\"{\i}ve first approximation that 
the duration of a tunnelling event is the barrier width divided by the 
velocity $\sqrt{2E/m}$.  Furthermore, it means that the wave vector
inside the forbidden region is imaginary, leading to exponential decay
(or growth), in lieu of the accumulation of phase.  As no phase is
accumulated (except near the boundaries, where the relative amplitudes
of the evanescent and anti-evanescent waves are varying 
significantly), the group delay becomes independent of barrier
thickness for ``opaque'' barriers ({\it i.e.}, barriers much wider
than an exponential decay length).  For wide enough barriers, a
wave packet peak may emerge from the far side at a time significantly
shorter than what would be required for a peak travelling at the
vacuum velocity of light $c$.  As long as the bandwidth of the signal
(typically taken to be gaussian in theoretical work, as well as in
certain experiments) is sufficiently small, an essentially undistorted
(but greatly attenuated) pulse appears to have travelled faster than
light.

As explained in \cite{Chiao=1997ProgOpt} and references therein, no 
{\it information} can be sent faster than light using these effects.
To sum the arguments up briefly, the wave exiting the barrier can be
expressed by integrating a {\it perfectly causal} response function:
\begin{equation}
\Psi(x=d,t) = \int_{0}^{\infty}d\tau f(\tau) \Psi(x=0,t-d/c-\tau) \; .
\end{equation}
The behaviour of $\Psi(x=0)$ at times $t_{0} > t-d/c$ has no impact
whatever on the behaviour of $\Psi(x=d,t)$.  Remarkably, the effect
of a tunnel barrier is to perform an extrapolation into the future
(a Taylor expansion\cite{Japha=1996,Diener=1996}) by combining 
paths with slightly
different delays $180^{\circ}$ out of phase with one 
another\cite{Steinberg=1993MOR,Steinberg=1995RECH}.  Whether for a gaussian wave packet 
or a piece by Mozart, the outcome is a convincing reproduction of the
incident wave form; some workers therefore argue that signal 
propagation for such frequency-band-limited waves indeed occurs
faster than $c$\cite{Nimtz=1997}.

Naturally, the difficulty lies with attempts to define (or to avoid
defining) ``information'' and ``signal.''  Only recently have workers
attempted to apply a rigorous definition \`{a} la 
Shannon\cite{Kurizki=1998}, but in our opinion, this application has 
not yet clarified matters.  The cases of superluminal propagation all 
occur when the ``advance'' of the transmitted wave remains within a 
region over which the function can be analytically continued.  It is 
well known that no {\it abrupt} disturbance would propagate faster 
than $c$, and recent electronic-circuit 
experiments\cite{Mitchell=1997} demonstrate that this is entirely 
consistent with the ability of superluminal systems to ``guess'' the 
future by analytic continuation-- the new feature is simply that at 
any point of nonanalyticity, the system will guess wrong.

Those who argue for superluminal signal velocities suggest that since 
bandwidths are always limited in practice, such abrupt fronts and
``points of nonanalyticity'' do not exist.  A signal defined 
practically, in terms of a well-defined waveform with a good 
signal-to-noise ratio, may indeed arrive sooner when tunnelling
than when propagating freely (although the time-integrated power
will always be less in the presence of the barrier).  The question
of how best to approximate physical reality-- with infinite tails
in the frequency domain and abrupt changes in time, or with infinite
``precursors'' in time and sharp cutoffs in frequency-- is an
interesting one, often sidestepped.  Alas, we choose for now to remain
within
this tradition.  In our opinion, when one talks of a signal velocity,
one implicitly refers to the velocity of {\it new} information arriving
at a receiver.  If one accepts that a transmitted waveform is strictly
band-limited in frequency, and can therefore be predicted well in 
advance of its peak by analytic continuation, then the peak does not
carry any new information.  When information is found at the receiver,
there is no reason to refer this back to the time of emission of the 
peak.  In particular, equation (1) shows that information about $x=0$
at times $t<-d/c$ will {\it always} suffice to predict what a receiver
observes at $x=d$ at times up to $t=0$.  If any new information is
sent after $t=-d/c$, the receiver will be unaware of it until a time
$t>0$.  The possibility of a {\it shape} appearing at $t<0$ and 
resembling a shape emitted at $t>-d/c$ in no way alters this.  It means
merely that the information available (even under free propagation) was
already sufficient for {\it predicting} the shape which was to come.  
(In the case of Mozart travelling less than a metre at several times 
the speed of light, this is another way of saying that it is possible 
to extrapolate a $10-20$ kHz signal a nanosecond into the future 
without significant distortion!)  In sum, a tunnel barrier acts as a 
marvelous analog computer for signal processing.  Under appropriate 
circumstances, it can take the leading edge of a signal and reshape 
it into a facsimile of the signal's main content.  While in 
conjunction with amplification\cite{Steinberg=1994DOUB}, this may even 
turn out to be useful, it is no different from what a digital signal 
processing chip could do by application of equation (1) to a 
freely-propagating waveform.  Therefore, as in the latter case, we 
conclude that tunnelling may reduce the delay necessary to {\it 
process} an incoming signal and {\it extract} information from it, 
but it in no 
way alters the arrival time of the new information itself. 

\section{Motivation: Interaction times and nonlocality}
It has long been recognized\cite{Buttiker=1982} that the delay of a 
wave packet peak need not be the unique definition of a tunnelling 
time (need not, in fact, have any physical significance at all 
in the view of those authors).  Many other times have been proposed to 
describe the ``interaction'' of a particle with a tunnel barrier, or 
the duration of its ``sojourn'' or ``dwell'' in the forbidden region.  
Using the weak-measurement formalism of Aharonov and 
Vaidman\cite{Aharonov=1988,Aharonov=1990}, we have found that many of 
these can be unified, and shown to correspond to a general 
description of von Neumann-style quantum measurements on a {\it 
subensemble} post-selected to contain only transmitted 
particles\cite{Steinberg=1995WK,Steinberg=1995PRA}.  The classic 
thought-experiment of this sort is the Larmor 
clock\cite{Buttiker=1983}, but application of the weak-measurement 
formalism makes it straightforward to obtain physical interpretations 
of the several components of the Larmor time, as well as their 
generalisation to a broader class of measurements.

As discussed in \cite{Steinberg=1998}, a number of these 
predictions are striking enough to merit experimental test.  
Summarized briefly, we predict that as defined by the outcome of a 
classically-conceived measurement, the time spent by a tunnelling 
particle near the middle of a barrier is exponentially small.  On the 
other hand, transmitted particles spend roughly equal amounts of time 
near the two classical turning points, while reflected particles only 
spend time near the entrance.  Throughout the entire region, attempts 
to detect the particle do on the other hand lead to quantum 
back-action, which is at the origin of the interpretational 
difficulties.

Perhaps more interesting, it appears that for these post-selected 
subensembles, there may be a measurable sense in which a particle can 
be in two places at one time.  Here we are not referring to the 
well-known fact that a wave function passes through {\it both slits} 
of a two-slit interferometer, or in general to the idea that every 
outcome is an integral over many Feynman paths.  In these cases, no 
attempt is made to {\it measure} the position of the particle at the 
slits.  Most discussion centers on how such a measurement, if 
sufficiently accurate to determine which slit is traversed, destroys 
interference.  We are interested in the reverse scenario.  We 
consider such a gentle, Larmor-style measurement that on no individual 
occasion can we tell whether or not a particle was observed.  By 
building up enough statistics, we can nevertheless draw conclusions 
about how much time the average particle spent interacting with the measuring 
device.  In the Larmor framework, the particle is a spin initially 
polarized along the $x$-direction, and the measuring apparatus is a 
weak magnetic field along $z$, confined to a small region of space.  
This field sets a Larmor precession frequency $\omega_{L}$, and a 
particle spending some length of time $t \ll 1/\omega_{L}$ in the 
interaction region will precess through an angle 
$\theta_{y}=\omega_{L}t$.  For a spin-1/2 particle, this angle is of 
course only measurable by averaging over many trials, but it is 
straightforward to separate transmitted and reflected particles, so as 
to measure their polarisations (and hence dwell times) separately.

We have made predictions for the outcome of such measurements, in the 
form of time-dependent ``conditional probability 
distributions''\cite{Steinberg=1995PRA}.  These distributions reflect 
the anomalously fast (potentially superluminal) arrival of a wave 
packet peak.  In fact, for thick enough barriers, one can choose 
spacelike-separated regions of spacetime centered at opposite edges 
of the barrier in such a way that the conditional probability of the 
particle being in either one is arbitrarily close to 100\%, since the
wave packet peaks themselves are spacelike-separated.  The question 
arises: can a single particle have an effect on two 
spacelike-separated measuring devices?  There is no problem with 
causality, since the {\it emission} of the particle may be in causal 
contact with both measuring devices.  Only the locality of a quantum 
particle is called into question, but in a way which cannot be 
directly addressed in familiar situations such as the two-slit 
experiment, the EPR effect, or the Aharonov-Bohm effect.

Consider a {\it Gedankenexperiment} along the lines illustrated in 
the spacetime diagram of Figure 1.  A wave packet tunnels superluminally
through a barrier wide enough that the bulk of the transmitted packet
is causally disconnected from the bulk of the incident packet.  
In this case, two weak-measurement devices could be turned on in such
a way that the incident particles have a probability arbitrarily close
to 100\% of interacting with the first, the transmitted particles have
a probability arbitrarily close to 100\% of interacting with the 
second, {\it and} that the two measurement regions are entirely 
spacelike separated.  The two measurement regions might, for example,
be pulsed magnetic fields pointing along {\bf +z} to the left of
the barrier and along {\bf -z} to the right of the barrier.  In this
way, a spin will have precessed in one direction if it sees only the
first measurement region, in the other direction if it sees only the
second, and not at all if it sees both for equal lengths of time.
After transmission, the particles are filtered
in energy in such a way that their time of emission can no longer
be determined, and it is impossible to trace a particle's trajectory
back to either region.

One could conceive of a model of reality which forbids any single particle 
from being affected measurably by two spacelike-separated devices.  
Most physicists seem to subscribe to such a model, and it is not yet 
known whether this is truly consistent with quantum mechanics.  Based
on the weak measurement picture, we expect that each particle will 
indeed be affected by both measuring devices.  In other words, if only 
one field or the other is switched on, the spins will precess.  But if 
both fields are on, rather than observing a mixture of spins which 
have precessed clockwise and spins which have precessed 
counter-clockwise, we believe that the spins will remain essentially
unaffected.  This is supported by calculations using the 
time-independent
Schr\"odinger equation, which show that to first order, the effects 
of two such Larmor fields cancel out identically.  We are currently 
planning to perform true time-dependent calculations.  We intend to 
construct inequalities constraining ``corpuscular'' models in which a 
particle cannot be affected by both fields, and to test these 
inequalities by studying tunnelling atoms.  This will constitute a 
test of single-particle nonlocality of a new sort, related to what 
Aharonov and coworkers term the ``reality of the wave 
function''\cite{Aharonov=1993AAV}.

\section{The atom optics experiment}

We believe that in order to study these quantum subensembles, 
laser-cooled atoms offer a unique tool.  They can routinely be 
cooled into the quantum regime, where their de Broglie wavelengths 
are on the order of microns, and their time evolution takes place in 
the millisecond regime.  They can be directly imaged, and if they are 
made to impinge on a laser-induced tunnel barrier, transmitted and 
reflected clouds should be spatially resolvable.  With various 
internal degrees of freedom (hyperfine structure as well as Zeeman 
sublevels), they offer a great deal of flexibility for studying the 
various interaction times and nonlocality-related issues.  In 
addition, extensions to dissipative interactions and questions related 
to irreversible measurements and the quantum-classical boundary are 
easy to envision.\cite{Steinberg=1998}

We are working on an atom-optics experiment which will
let us directly test these questions.  We  start with a 
sample of laser-cooled Rubidium atoms transferred from a magneto-optic
trap (MOT) and optical molasses cooling phase to a magnetic trap at
approximately 6 $\mu K$.  We plan to use a
tightly focussed beam of intense light detuned far to the blue of the 
D2 line to create a dipole-force potential for the 
atoms\cite{Rolston=1992,Miller=1993FORT,Davidson=1995}.  Using a 500 
mW 
laser at 770 nm, we will be able to make repulsive potentials with 
maxima on the order of the Doppler temperature of the Rubidium vapour.
Acousto-optical modulation of the beam will let us shape these 
potentials with nearly total freedom, such that we can have the atoms 
impinge on a thin plane of repulsive light, whose width would be on 
the order of the cold atoms' de Broglie wavelength.  This is because 
the beam may be focussed down to a spot several microns across 
(somewhat larger than the wavelength of atoms in a MOT, but of the 
order of that of atoms just below the recoil temperature, and hence
accessible by a combination of cooling and selection techniques).  
This focus may be rapidly displaced \cite{Steinberg=ICAP1996,Rudy=1997} by 
using acousto-optic modulators.  As the atomic motion is in the 
mm/sec range, the atoms respond only to the time-averaged intensity, 
which can be arranged to have a nearly arbitrary profile.

As a second stage of cooling, we follow the MOT and optical molasses
with an improved variant of a proposal termed ``delta-kick 
cooling''\cite{Ammann=1997}.  In our version, the millimetre-sized 
cloud is allowed to expand for about ten milliseconds, to several times 
its initial size.  This allows individual atoms' positions $x_{i}$
 to become strongly correlated 
with atomic velocity, $x_{i} \approx v_{i}t_{\rm free}$.  Magnetic 
field coils are then used in either a quadrupole or a harmonic 
configuration to provide a position-dependent restoring force for
a short period of time.  By 
proper choice of this impulse, one can greatly reduce the rms velocity
of the atoms.  Figure 2 shows free-expansion data comparing atoms 
expanding from an optical molasses with and without a quadrupole 
delta-kick.  As can be seen by the narrow horizontal width of the 
latter distribution at late times (the narrow peak is difficult to 
distinguish from these pictures at early times, but already present
in the raw data), the thermal velocity has been greatly reduced.  The 
one-dimensional temperature appears to be about 700 nK, corresponding 
to a de Broglie wavelength of about half a micron.  We are currently 
working on improving this temperature by producing stronger, more 
harmonic potentials, and simultaneously providing an antigravity potential 
 in order to increase the interaction time.

However, the tunnelling probability through a 5-micron focus will 
still be negligible at these temperatures.  Furthermore, the 
exponential dependence of the tunnelling rate on barrier height will 
be difficult to distinguish from the exponential tail of a thermal 
distribution at high energies.  We will therefore follow the 
delta-kick with a velocity-selection phase\cite{Steinberg=1998}.  
By using the same beam which is to form a tunnel barrier, but 
increasing the width to many microns, we will be able to ``sweep'' 
the lowest-energy atoms from the center of the magnetic trap off to the 
side, leaving the hotter atoms behind.  Figure 3 shows the results 
of simulations for atoms at an initial temperature of 900 nK, trapped
in a 5 G/cm field (corresponding to 300 $\mu$K/cm).  Superposing a 
20-$\mu$m gaussian beam with a peak height of 300 nK onto this 
V-shaped potential creates a local minimum with a depth of 70 nK, 
supporting a single quasi-bound state with a kinetic energy of 13 
nK.  Translating this barrier through the cloud at $0.5$ mm/s is seen 
to transfer about 7\% of the atoms into this ground state.  This new, 
smaller sample will have a thermal de Broglie wavelength of 
approximately $3.5 \mu$m, leading to a significant tunnelling 
probability through a 10-micron barrier.  We expect rates on 
the order of 1\% per secular period, causing the auxiliary trap to 
decay via tunnelling on a timescale of the order of 100 ms. 

Once these cooling and selection techniques are perfected, we will 
have a unique system in which to study tunnelling.  By using optical 
pumping, stimulated Raman transitions, and other such probes, we will 
be able to go beyond simple wave packet studies to investigate the 
interactions of tunnelling atoms while in the forbidden region 
itself.  In this way, we hope to shed new light on this fascinating 
phenomenon, but also on nonlocality in quantum mechanics more 
generally.

\newpage

{\bf Figure Captions}

Fig. 1.  As explained in the text, this {\it Gedankenexperiment} is 
intended to examine whether a single tunnelling particle can be 
simultaneously affected by two spacelike-separated ``weak-measurement
devices,'' for example, pulsed magnetic fields.  The superluminally
transmitted particle is passed through an energy-filter which erases
all information about its time of emission, and the precession angle 
of its spin should provide information about which of the interaction 
regions it traversed.

Fig. 2.  These fluorescence images show the expansion of a cloud of 
ultracold Rubidium atoms over a 27 ms period.  The isotropic expansion
in the upper sequence, directly from an optical molasses, indicates a 
temperature of approximately 9 $\mu$K.  The lower sequence shows an
analogous expansion after a ``quadrupole kick'' is used to further cool
the atoms one-dimensionally.  The dark central stripe visible in the
final frames has been analyzed, and corresponds to a temperature of
approximately 700 nK.

Fig. 3.  This quantum-mechanical simulation demonstrates what we expect
to achieve with our dipole-force velocity-selection.  Starting with an
atom cloud near 1 $\mu$K and sweeping an appropriately tuned 20 $\mu$m
laser beam through the atoms adiabatically, we will create a very small
auxiliary potential well.  The probability of transfer exhibits steps
as a function of well depth, indicating the number of quasi-bound 
states supported.  A well with a single bound state is seen to capture
about 7 \% of the atoms, in a state with a kinetic temperature of only
13 nK, corresponding to a de Broglie wavelength of about 3.5 $\mu$m.  
Such a state will be an ideal source for our atom-tunnelling 
experiments.

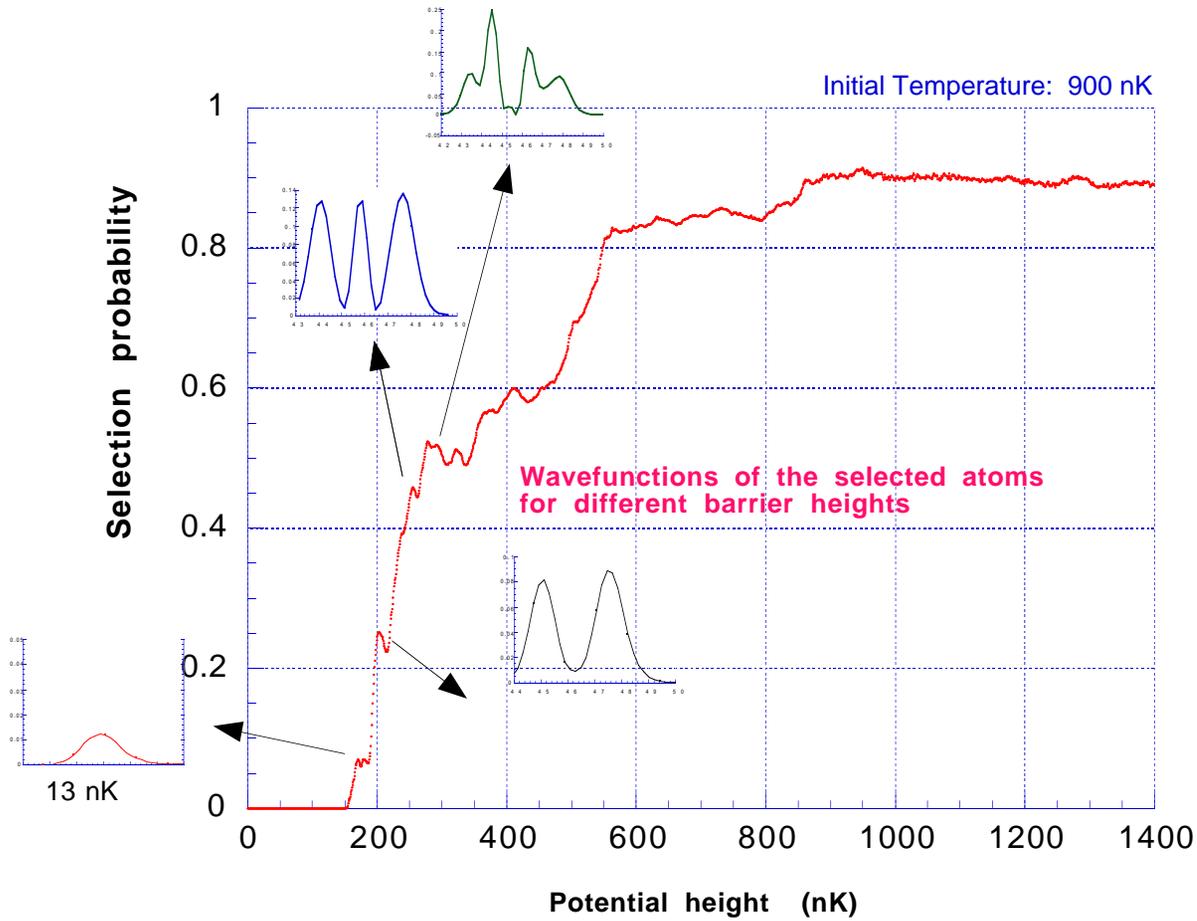
\end{document}